\documentclass[pre,tighten]{revtex4}
\usepackage[latin1]{inputenc}
\usepackage{amssymb,amsmath}
\usepackage{graphicx}% Include figure files
\begin{document}
\title{Hydrodynamic equations for granular mixtures}
\author{Vicente Garz\'{o}}
\email[E-mail: ]{vicenteg@unex.es}
\address{Departamento de F\'{\i}sica, Universidad de Extremadura, E-06071
Badajoz, Spain}
\author{James W. Dufty}
\email[E-mail: ]{dufty@phys.ufl.edu}
\address{Department of Physics, University of Florida, Gainesville, Florida 32611}

\begin{abstract} 

Many features of granular media can be modeled by a fluid of hard spheres
with inelastic collisions. Under rapid flow conditions, the macroscopic
behavior of grains can be described through hydrodynamic equations
accounting for dissipation among the interacting particles. A basis for the
derivation of hydrodynamic equations and explicit expressions appearing in
them is provided by the Boltzmann kinetic theory conveniently modified to
account for inelastic binary collisions. The goal of this review is to 
derive the hydrodynamic equations for a binary mixture of smooth
inelastic hard spheres. A normal solution to the Boltzmann equation 
is obtained via the Chapman-Enskog method for states near the {\em local}
homogeneous cooling state. The mass, heat, and momentum fluxes are obtained
to first order in the spatial gradients of the hydrodynamic fields, and the
set of transport coefficients are determined in terms of the restitution
coefficients and the ratios of mass, concentration, and particle sizes. 
As an example of their application, the dispersion relations for the
hydrodynamic equations linearized about a homogeneous state are obtained
and the conditions for stability are identified as functions of the wave
vector, the dissipation, and the parameters of the mixture. The
analysis shows that the homogeneous reference state is unstable to long
enough wavelength perturbations and consequently becomes inhomogeneous for
long times. The relationship of this instability to the validity of
hydrodynamics is discussed.

\end{abstract} 

\draft
\date{\today}
\maketitle

\section{Introduction}
\label{sec1}

Granular media under rapid flow conditions exhibit a great similarity to molecular fluids. This fact has stimulated the use of hydrodynamic-like type equations to describe the macroscopic behavior of such systems. The main difference from ordinary fluids is the absence of energy conservation, leading to both obvious and subtle modifications of the Navier-Stokes hydrodynamic equations. To isolate the effects of such collisional dissipation from other important properties of granular media, an idealized microscopic model system is usually considered: a system composed by smooth hard spheres with inelastic collisions. As in the elastic case, the collisions are  specified in terms of the change in relative velocity at contact but with a decrease in the magnitude of the normal component measured by a positive restitution coefficient $\alpha \leq 1$. This parameter distinguishes the ideal granular fluid ($\alpha <1$) from the ideal normal fluid ($\alpha=1$). 

Although many efforts have been devoted in the past few years in the
understanding of granular fluids, the derivation of the form of the
transport coefficients is still a subject of interest and controversy. In
the low-density regime, this problem can be addressed by using the Boltzmann
kinetic theory conveniently modified to account for inelastic collisions\cite
{BDS97}. The conditions for hydrodynamics are expected to be similar
to those for \ normal fluids. For a given initial state there are two stages
of evolution. First, during the kinetic stage there is rapid velocity
relaxation to a ``universal'' velocity distribution that depends on the
average local density, temperature, and flow velocity. Subsequently, the
hydrodynamic stage is described through a slower evolution of these local
hydrodynamic fields as they approach uniformity. The solution to the
Boltzmann equation in this second stage is said to be ``normal'', where all
space and time dependence of the distribution function occurs through the
hydrodynamic fields. The Chapman-Enskog method \cite{CC70} provides a means
to construct explicitly the form of this normal solution as a perturbation
expansion in the spatial gradients of the fields. This solution is then used
to evaluate the fluxes in the macroscopic balance equations in terms of
these gradients. To lowest order the balance equations become the granular
Euler equations; to second order they are the granular Navier-Stokes
equations. In carrying out this analysis, explicit forms for the transport
coefficients are obtained as functions of the restitution coefficient and
other parameters of the collision operator. This derivation of hydrodynamics
from the Boltzmann equation has been widely covered in the case of a
monocomponent gas where the particles are of the same mass and size \cite
{BDKS98}. However, a real granular system is generally characterized by some degrees of polidispersity in density and size, which leads to phenomena very often observed in nature and experiments, such as separation or segregation. Several attempts to apply the Boltzmann equation to derive transport coefficients for a multicomponent system began time ago \cite{Jenkins}, but the technical difficulties of the analysis entailed approximations that limited their accuracy. In addition, all these works are based on the assumption of energy equipartition so that the partial temperatures $T_i$ are made equal to the global granular temperature $T$. Nevertheless, kinetic theory studies\cite{GD99,BT02,MP02,NK02}, computer simulations\cite{MG02,DHGD02,CH02} and even real experiments \cite{WP02,FM02} have clearly shown the breakdown of energy equipartition. As a consequence, many of the previous results obtained for granular mixtures must be reexamined by using a kinetic theory which takes into account the effect of  temperature differences on the transport coefficients. 

The goal of this paper is to derive the hydrodynamic equations of a granular binary mixture from the Boltzmann kinetic theory. These equations are derived by applying the Chapman-Enskog expansion about a {\em local} homogeneous cooling state (LHCS) that is analogous to the local equilibrium state for a gas with elastic collisions. In the first order of the expansion, the irreversible parts of the mass, heat, and momentum fluxes are calculated and the eight transport coefficients identified. These coefficients are expressed in terms of the solutions to a set of coupled linear integral equations\cite{GD02}. The analysis carried out here is more complete than previous studies \cite{Jenkins}; it is exact to leading order in the dissipation but not limited to weak dissipation. However, for practical purposes, the above integral equations are solved approximately by using the leading terms in a Sonine polynomial expansion. Such approach  compares quite well with the results obtained from numerical solutions \cite{MG02,MG03} of the Boltzmann equation by means of the Direct Simulation Monte Carlo (DSMC) method\cite{B94}. The explicit knowledge of the transport coefficients allows
quantitative application of the nonlinear hydrodynamic equations to a number
of interesting problems for mixtures, such as segregation and separation.
Here, the simplest example of an application is considered, small
perturbations of a spatially homogeneous state. The dispersion relations for
the hydrodynamic modes are obtained from the hydrodynamic equations
linearized about the homogeneous form of the LHCS. Linear stability analysis
shows two shear modes and a heat mode to be unstable for long wavelength
excitations. This means that small perturbations or fluctuations about the
homogeneous state will grow. The implications of this for the derivation of
the hydrodynamic equations are discussed here as well.

The plan of the paper is as follows. In Sec.\ \ref{sec2}, we review the 
Boltzmann equation and associated macroscopic conservation laws. We also give a brief 
survey of the application of the Chapman-Enskog method around the LHCS. This state is analyzed in Section \ref{sec3} while in Sec.\ \ref{sec4} we give the form of the Navier-Stokes hydrodynamic equations. Theoretical results are compared with simulation data for the temperature ratio and the shear viscosity coefficient.  Section \ref{sec5} deals with an analysis of the stability of the linearized hydrodynamic equations. We close the paper in Section \ref{sec6} with a discussion of the results presented.

\section{Boltzmann equation and conservation laws}
\label{sec2}

We consider a binary mixture of smooth hard spheres of 
masses $m_{1}$ and $m_{2} $, and diameters $\sigma _{1}$ and $\sigma _{2}$. The inelasticity of 
collisions among all pairs is characterized by three independent constant 
coefficients of normal restitution $\alpha_{11}$, $\alpha_{22}$, and 
$ \alpha_{12}=\alpha_{21}$, where $\alpha_{ij}$ is the restitution 
coefficient for collisions between particles of species $i$ and $j$. In the low-density regime, the 
distribution functions $f_{i}({\bf r},{\bf v};t)$ $
(i=1,2)$ for the two species are determined from the set of nonlinear 
Boltzmann equations \cite{BDS97}
\begin{equation} 
\left( \partial _{t}+{\bf v}_{1}\cdot \nabla \right) 
f_{i}=\sum_{j}J_{ij}\left[ {\bf v}_{1}|f_{i}(t),f_{j}(t)\right] \;,
\label{2.1} 
\end{equation} 
where  the Boltzmann collision operator $J_{ij}\left[ {\bf v}_{1}|f_{i},f_{j}\right] $ describing 
the scattering of pairs of particles is  
\begin{eqnarray} 
J_{ij}\left[ {\bf v}_{1}|f_{i},f_{j}\right] &=&\sigma _{ij}^{2}\int d{\bf v} 
_{2}\int d\widehat{\boldsymbol {\sigma }}\,\Theta (\widehat{{\boldsymbol {\sigma }}} 
\cdot {\bf g}_{12})(\widehat{\boldsymbol {\sigma }}\cdot {\bf g}_{12})  \nonumber 
\\ &&\times \left[ \alpha _{ij}^{-2}f_{i}({\bf r},{\bf v}_{1}^{\prime 
},t)f_{j}( {\bf r},{\bf v}_{2}^{\prime },t)-f_{i}({\bf r},{\bf v}
_{1},t)f_{j}({\bf r}, {\bf v}_{2},t)\right] \;,  \label{2.2} 
\end{eqnarray} 
where $\sigma _{ij}=\left( \sigma _{i}+\sigma _{j}\right) /2$, $\widehat{
\boldsymbol {\sigma}}$ is a unit vector along their line of centers, $\Theta $ is 
the Heaviside step function, and ${\bf g}_{12}={\bf v}_{1}-{\bf v}_{2}$. The 
primes on the velocities denote the initial values $\{{\bf v}_{1}^{\prime},  
{\bf v}_{2}^{\prime }\}$ that lead to $\{{\bf v}_{1},{\bf v}_{2}\}$ 
following a binary collision:  
\begin{equation} 
{\bf v}_{1}^{\prime }={\bf v}_{1}-\mu _{ji}\left( 1+\alpha _{ij}^{-1}\right) 
(\widehat{{\boldsymbol {\sigma }}}\cdot {\bf g}_{12})\widehat{{\boldsymbol {\sigma }}} 
,\quad {\bf v}_{2}^{\prime }={\bf v}_{2}+\mu _{ij}\left( 1+\alpha 
_{ij}^{-1}\right) (\widehat{{\boldsymbol {\sigma }}}\cdot {\bf g}_{12})\widehat{ 
\boldsymbol {\sigma}} , \label{2.3} 
\end{equation} 
where $\mu _{ij}=m_{i}/\left( m_{i}+m_{j}\right)$.

The relevant hydrodynamic fields are the number densities $n_{i}$, 
the flow velocity ${\bf u}$, and the ``granular'' temperature $T$. 
They are defined in terms of moments of the distributions $f_{i}$ as  
\begin{equation} 
n_{i}=\int d{\bf v}f_{i}({\bf v})\;,\quad \rho {\bf u}=\sum_{i}\int 
d {\bf v}m_{i}{\bf v}f_{i}({\bf v})\;,  \label{2.4} 
\end{equation} 
\begin{equation} 
nT=p=\sum_{i}\int d{\bf v}\frac{m_{i}}{3}V^{2}f_{i}({\bf v})\;, 
\label{2.5} 
\end{equation} 
where $n=n_{1}+n_{2}$ is the total number density, $\rho 
=m_{1}n_{1}+m_{2}n_{2}$ is the total mass density, $p$ is the 
hydrostatic pressure, and ${\bf V}={\bf v}-{\bf u}$ is the peculiar velocity. At a kinetic level, it is convenient to introduce the 
kinetic temperatures $T_i$ for each species defined as 
\begin{equation}
\label{2.6}
\frac{3}{2}n_iT_i=\int d{\bf v}\frac{m_{i}}{2}V^{2}f_{i}.
\end{equation}

The collision operators conserve the particle number of each species and the 
total momentum but the total energy is not conserved:
\begin{equation}
\label{2.7}
\int d{\bf v}J_{ij}[{\bf v}|f_i,f_j]=0,\quad 
\sum_{i,j}\int d{\bf v}m_i{\bf v}J_{ij}[{\bf v}|f_i,f_j]=0,
\end{equation}
\begin{equation}
\label{2.8}
\sum_{i,j}\int d{\bf v}\case{1}{2}m_iV^2J_{ij}[{\bf 
v}|f_i,f_j]=-\case{3}{2}nT\zeta,
\end{equation}
where $\zeta$ is identified as the {\em cooling rate} due to inelastic 
collisions among all species. The macroscopic balance equations follow 
from the Boltzmann equation (\ref{2.1}) and Eqs.\ (\ref{2.7}) and 
(\ref{2.8}). They are given by  
\begin{equation} 
D_{t}n_{i}+n_{i}\nabla \cdot {\bf u}+\frac{\nabla \cdot {\bf j}_{i}}{m_{i}} 
=0\;,  \label{2.9} 
\end{equation} 
\begin{equation} 
D_{t}{\bf u}+\rho ^{-1}\nabla {\sf P}=0\;,  \label{2.10} 
\end{equation} 
\begin{equation} 
D_{t}T-\frac{T}{n}\sum_{i}\frac{\nabla \cdot {\bf j}_{i}}{m_{i}}+\frac{2}{3n} 
\left( \nabla \cdot {\bf q}+{\sf P}:\nabla {\bf u}\right) 
=-\zeta T\;. \label{2.11} 
\end{equation} 
In the above equations, $D_{t}=\partial _{t}+{\bf u}\cdot \nabla $ is the 
material derivative,  
\begin{equation} 
{\bf j}_{i}=m_{i}\int d{\bf v}\,{\bf V}\,f_{i}({\bf v})
\label{2.12} 
\end{equation} 
is the mass flux for species $i$ relative to the local flow,  
\begin{equation} 
{\sf P}=\sum_{i}\,\int d{\bf v}\,m_{i}{\bf V}{\bf V}\,f_{i}({\bf  
v})  \label{2.13} 
\end{equation} 
is the total pressure tensor, and  
\begin{equation} 
{\bf q}=\sum_{i}\,\int d{\bf v}\,\case{1}{2}m_{i}V^{2}{\bf V} 
\,f_{i}({\bf v})  \label{2.14} 
\end{equation} 
is the total heat flux. 

The utility of the balance equations (\ref{2.9})--(\ref{2.11}) is limited without further specification of the fluxes and the cooling rate, which in general have a complex dependence on space and time. However, for sufficiently large space and time scales, one expects that the system reaches a hydrodynamic regime in which all the space and time dependence is given entirely through a functional dependence on the hydrodynamic fields $n_i$, ${\bf u}$, and $T$. The corresponding functional dependence of ${\bf j}_i$, ${\sf P}$, and ${\bf q}$ on these fields are called constitutive equations and define the transport coefficients of the mixture. The primary feature of a hydrodynamic description is the reduction of the description from many microscopic degrees of freedom to a set of equations involving only  five local fields. At a kinetic level, the constitutive equations are obtained when one admits the existence of a {\em normal} solution to the Boltzmann equation where the velocity distribution function depends on ${\bf r}$ and $t$ only through its functional dependence on the fields, namely,  
\begin{equation}
\label{2.15}
f_i({\bf r}, {\bf v}_1,t)=f_i[{\bf v}_1|n_1({\bf r},t), T({\bf r},t), {\bf u}({\bf r},t)].
\end{equation}  
The Chapman-Enskog method \cite{CC70} generates this normal solution explicitly by means of 
an expansion in gradients of the fields:
\begin{equation} 
f_{i}=f_{i}^{(0)}+\epsilon \,f_{i}^{(1)}+\epsilon^2 \,f_{i}^{(2)}+\cdots \;, 
\label{2.16} 
\end{equation} 
where $\epsilon$ is a formal parameter measuring the nonuniformity of 
the system. The local reference state $f_i^{(0)}$ is chosen to give 
the same first moments as the exact distribution $f_i$. The time derivatives 
of the fields are also expanded as $\partial_t=\partial_t^{(0)}+\epsilon 
\partial_t^{(1)}+\cdots$. The coefficients of the time derivative expansion 
are identified from the balance equations (\ref{2.9})--(\ref{2.11}) after 
expanding the fluxes, and the cooling rate $\zeta$ in a similar series as 
(\ref{2.16}). 
We close this section with some comments about the partial
temperatures $T_{i}$ introduced in Eq.\ (\ref{2.6}). In contrast to the
global temperature $T$ these partial temperatures are not
hydrodynamic fields and do not appear directly in the balance equations (\ref
{2.9})--(\ref{2.11}) and in the normal solution (\ref{2.15}). However, as
indicated in the next section, these partial temperatures characterize
differences in the velocity distributions for the different species and
therefore are important for their quantitative description. Ultimately, the
transport coefficients obtained depend on the ratio $T_{1}/T_{2}$
which is expressed in terms of the mechanical properties of the particle
collisions, through the requirement that the solution be normal. 

\section{Local homogeneous cooling state}
\label{sec3}

To zeroth order in $\epsilon $, the kinetic equations (\ref{2.1}) become  
\begin{equation} 
\partial_{t}^{(0)}f_{i}^{(0)}=\sum_{j}\,J_{ij}[f_{i}^{(0)},f_{j}^{(0}]\;.
\label{3.1} 
\end{equation} 
The normal form of $f_i^{(0)}$ requires that all its time dependence occurs only $n_i$, ${\bf u}$, and $T$. Consequently, $f_i^{(0)}$ must be of the form 
\begin{equation} 
f_{i}^{(0)}({\bf v})=n_{i}v_{0}^{-3}\Phi _{i}\left(V/v_{0}\right) \;, 
\label{3.4} 
\end{equation} 
where $v_{0}^{2}(t)=2T(t)(m_{1}+m_{2})/\left( m_{1}m_{2}\right) $ is a thermal velocity defined in terms of the temperature $T(t)$ of the mixture. The balance equations (\ref{2.9})--(\ref{2.11}) to this order become $\partial _{t}^{(0)}x_{i}=\partial_t^{(0)} {\bf u}=0, T^{-1}\partial 
_{t}^{(0)}T=-\zeta ^{(0)}$. Here, $x_i=n_i/n$ is the mole fraction of species $i$ and the cooling rate $\zeta ^{(0)}$ is determined by Eq.\ (\ref{2.8}) to zeroth order.  Hence, all the time dependence to this order is through the temperature and so, in dimensionless form, Eq.\ (\ref{3.1}) becomes the following time independent equation for $\Phi_i$:  
\begin{equation} 
\frac{1}{2}\zeta ^*\frac{\partial}{\partial {\bf v}^*_1}\cdot \left( {\bf V}^*
_{1}\Phi_i\right) =\sum_{j}\,J^*_{ij}[\Phi_i,\Phi_j]\;, 
\label{3.5} 
\end{equation} 
where ${\bf V}^*={\bf V}/v_0$, $\zeta^*=\zeta_i/n\sigma_{12}^2v_0$, and $J^*_{ij}=v_0^2J_{ij}/nn_i\sigma_{12}^2$. 
Since the distribution functions are isotropic, it follows from Eqs.\ (\ref 
{2.12})--(\ref{2.13}) that the zeroth order mass and heat fluxes 
vanish while, for the same reason, the momentum flux is diagonal with a 
coefficient that is just the sum of the partial pressures, i.e.,  
\begin{equation} 
{\bf j}_{i}^{(0)}={\bf 0,}\,\hspace{0.3in}{\bf q}^{(0)}={\bf 0,}\,\hspace{
0.3in}P_{\alpha \beta }^{(0)}=p\delta _{\alpha \beta }.  \label{3.7} 
\end{equation} 
In summary, Eq.\ (\ref{3.4}) provides the normal solution to lowest order
in the gradients and use of Eq.\ (\ref{3.7}) in the balance equations gives
the corresponding Euler level hydrodynamics. Note that $\zeta ^{(0)}$
is determined self-consistently in the solution to Eq.\ (\ref{3.5}).

Just as for the one component gas case\cite{GS95,NE98}, the exact form of $\Phi_i$ has not yet been found, although a good approximation for thermal velocities can be obtained from an expansion in Sonine polynomials\cite{GD99}. In the leading order, $\Phi_i$ is given 
by 
\begin{equation}
\label{3.8}
\Phi_i(V^*)\to \left(\frac{\theta_i}{\pi}\right)^{3/2}e^{-\theta_iV^{*2}}
\left[1+\frac{c_i}{4}\left(\theta_iV^{*4}-5\theta_iV^{*2}+\frac{15}{4}\right)
\right].
\end{equation}
Here, $\theta _{i}=(m_{i}/\gamma _{i})\sum_{j}m_{j}^{-1}$ and 
$\gamma _{i}=T_{i}/T$.The coefficients $c_{i}$ (which measure 
the deviation of $\Phi _{i}$ from the reference Maxwellian) are determined consistently from the Boltzmann equation. An important observation is that the partial temperatures have been
introduced here since they provide the actual mean kinetic energies of the
distributions through Eq.\ (\ref{2.6}).  This would appear to contradict the
assumption of a \ normal solution by introducing additional variables in the
distribution functions. To see that this is not the case, the time evolution
for the temperature ratio $\gamma =T_{1}/T_{2}$  is calculated to be
\begin{equation}
\label{3.2}
\partial_t^{(0)} \ln \gamma=\zeta_2^{(0)}-\zeta_1^{(0)}
\end{equation}
where we have introduced the cooling rates $\zeta_i^{(0)}$ for the partial temperatures $T_i$ as 
\begin{equation}
\label{3.3}
\zeta_i^{(0)}=-\frac{2}{3n_iT_i}\sum_{j}\int d{\bf v}\case{1}{2}m_iV^2J_{ij}[{\bf 
v}|f_i^{(0)},f_j^{(0)}].
\end{equation}
The total cooling rate $\zeta^{(0)}$ can be expressed in terms of the coolings $\zeta_i^{(0)}$ as 
\begin{equation}
\label{3.3bis}
\zeta^{(0)}=T^{-1}\sum_i x_iT_i\zeta_i^{(0)}.
\end{equation}
 \begin{figure}
\includegraphics[width=0.5 \columnwidth]{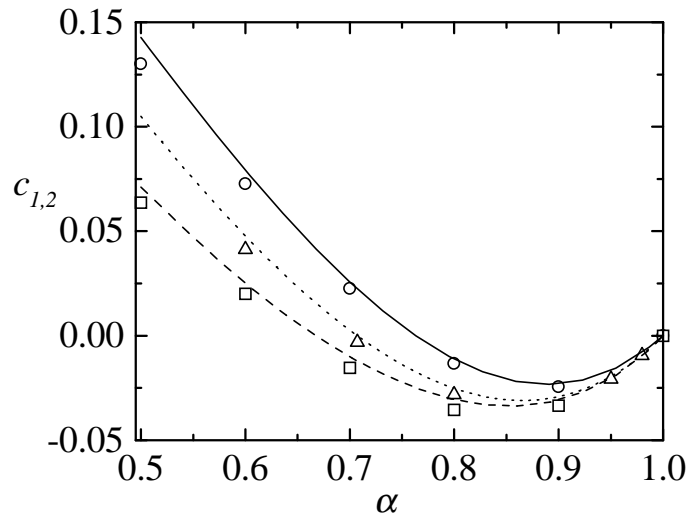}
\caption{Plot of the coefficients $c_i$ versus $\alpha$ for $\sigma_1/\sigma_2=1$, $x_1=\case{1}{2}$ and $m_1/m_2=2$. The solid line and the circles refer to $c_1$ while the dashed line and the squares correspond to $c_2$. The dotted line and the triangles refer to the common value in the monocomponent case. The lines are the theoretical results and the symbols are the simulation results.  
\label{fig1}}
\end{figure}
The fact that $f_i$ depends on time only through $T(t)$ necessarily implies that the temperature ratio $\gamma$ must be independent of time and so, Eq.\ (\ref{3.2}) gives the condition $\zeta_1^{(0)}=\zeta_2^{(0)}=\zeta^{(0)}$. In the elastic case, where $f_i^{(0)}$ is a local Maxwellian, the above condition yields $T_1(t)=T_2(t)=T(t)$ and the energy equipartition applies. However, in the inelastic case, the equality of the cooling rates leads to different values for the partial temperatures, even if one considers the Maxwellian approximation to $f_i^{(0)}$. 
Nevertheless, the constancy of $\gamma $ assures that the time
dependence of the distribution is entirely through $T(t)$, and in
fact the condition of equal cooling rates assures that the partial
temperatures can be expressed in terms of $T$.

\begin{figure}
\includegraphics[width=0.5 \columnwidth]{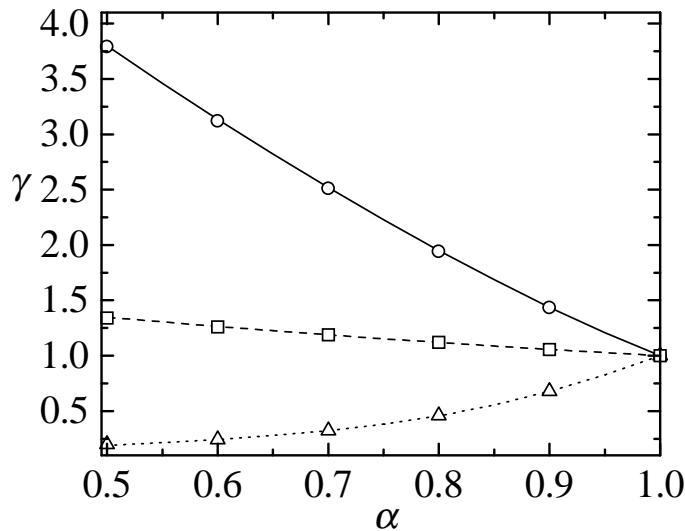}
\caption{Plot of the temperature ratio $\gamma\equiv T_1/T_2$ versus $\alpha$ for   $\sigma_1/\sigma_2=1$, $x_1=\case{2}{3}$ and three different values of the mass ratio: $m_1/m_2=0.1$ (dotted line and triangles), $m_1/m_2=2$ (dashed line and squares), and $m_1/m_2=10$ (solid line and circles). The lines are the theoretical results and the symbols are the simulation results.  
\label{fig2}}
\end{figure}

The explicit form of the approximation (\ref{3.8}) provides detailed predictions for 
the temperature ratio $T_1/T_2$ (through calculation of the cooling rates) and for the cumulants $c_i$ [from Eq.\ (\ref{3.5})] as functions of the mass ratio, size ratio, composition and restitution coefficients\cite{GD99}.  This completely fixes the approximate distribution function. In Fig.\ \ref{fig1}, we show the dependence of $c_1$ and $c_2$ on the (common) restitution coefficient $\alpha_{ij}\equiv \alpha$ for $\sigma_1/\sigma_2=1$, $x_1=\case{1}{2}$, and $m_1/m_2=2$. We also present the corresponding results obtained from a numerical solution \cite{MG02} of the Boltzmann equation by means of the DSMC method\cite{B94}. The agreement between the simulation data and the theoretical results is excellent. Further, the small values of the coefficients $c_i$  supports the assumption of a low-order truncation in the polynomial expansion of the distribution functions. 
One of the main findings of our theory is that, except for mechanically equivalent particles,  the partial temperatures are in general different ($\gamma\equiv T_1/T_2\neq 1$). This conclusion contrasts with all previous results\cite{Jenkins}  derived for granular mixtures, where it was implicitly assumed the equipartition of granular energy between both species (i.e., $\gamma=1$). The breakdown of the energy equipartition has been confirmed by computer simulations\cite{BT02,DHGD02,MP02,CH02} and by real experiments of vibrated mixtures in two \cite{WP02} and three \cite{FM02} dimensions. For the sake of illustration, in Fig.\ \ref{fig2} we plot the temperature ratio $\gamma$ versus $\alpha$ for $\sigma_1/\sigma_2=1$, $x_1=\case{2}{3}$, and several values of the mass ratio as given by the first Sonine approximation and by Monte Carlo simulations. The comparison between theory and simulation data shows an excellent agreement, even for strong dissipation. We also observe that for large differences in the mas ratio, the temperature differences are quite important. 

\section{Navier-Stokes hydrodynamic equations}
\label{sec4}

Implementation of the Chapman-Enskog method to the first order in $\epsilon$ has been carried out and without approximation recently\cite{GD02}. Technical details of these calculations will not be repeated here and only the main results are offered. The solution of the Boltzmann equation to first order in the spatial gradients can be written as 
\begin{equation}
\label{4.1}
f_i^{(1)}={\boldsymbol{\cal A}}_i\cdot \nabla x_1+{\boldsymbol{\cal B}}_i\cdot \nabla p+{\boldsymbol{\cal C}}_i\cdot \nabla T+{\cal D}_{i,k\ell}\nabla_{k}u_{\ell}.
\end{equation}
The coefficients ${\boldsymbol{\cal A}}_i$, ${\boldsymbol{\cal B}}_i$, ${\boldsymbol{\cal C}}_i$, and ${\cal D}_{i,k\ell}$ are functions of the peculiar velocity and the hydrodynamic fields.  These coefficients obey inhomogeneous integral equations involving the linearized Boltzmann operators. Solubility conditions for the existence of solutions has been also proven.  The corresponding constitutive equations found to this order are 
\begin{equation}
\label{4.2}
{\bf j}_1^{(1)}=-\frac{m_1m_2n}{\rho}D\nabla x_1-\frac{\rho}{p}D_p\nabla p-
\frac{\rho}{T}D'\nabla T,\quad {\bf j}_2^{(1)}=-{\bf j}_1^{(1)},
\end{equation}
\begin{equation}
\label{4.3}
{\bf q}^{(1)}=-T^2D''\nabla x_1-L\nabla p-\lambda\nabla T,
\end{equation}
\begin{equation}
\label{4.4}
P_{k\ell}^{(1)}=-\eta\left(\nabla_\ell u_k+
\nabla_k u_\ell-\frac{2}{3}\delta_{k\ell}\nabla \cdot {\bf 
u}\right).
\end{equation}
Furthermore, there is not contribution to the cooling rate at this order, i.e., $\zeta^{(1)}=0$. This is a consequence of the symmetry of the distribution function $f_1^{(1)}$ since the latter does not contain any contribution proportional to the divergence of the flow velocity field.  This property is special to the low density Boltzmann kinetic theory and such terms occur at higher densities \cite{GD99b}. The transport coefficients appearing in Eqs.\ (\ref{4.2})--(\ref{4.4}) are the diffusion coefficient $D$, the thermal diffusion coefficient $D'$, the pressure diffusion coefficient $D_p$, the 
Dufour coefficient $D''$, the thermal conductivity $\lambda$, the pressure 
energy coefficient $L$, and the shear viscosity $\eta$. These transport coefficients can be easily written in terms of ${\boldsymbol{\cal A}}_i$, ${\boldsymbol{\cal B}}_i$, ${\boldsymbol{\cal C}}_i$, and ${\cal D}_{i,k\ell}$. As in the elastic case\cite{CC70}, an accurate estimate of the transport coefficients can be obtained by using the leading terms in a Sonine polynomial expansion. Thus, the transport coefficients are known as explicit functions of the restitution coefficients and the parameters of the mixture (masses, sizes, and composition) and their expressions are not limited to weak inelasticity.

These expressions for the mass flux, the pressure tensor, the heat
flux, and the cooling rate provide the necessary constitutive equations to
convert the balance equations (\ref{2.9})--(\ref{2.11}) into a 
closed set of six independent equations for the hydrodynamic fields. Since Eqs.\ (\ref{4.2})--(\ref{4.4}) have been represented in terms of the gradients of the mole fraction 
$x_{1}$, the pressure $p$, the temperature $T$, and the flow velocity 
${\bf u}$, it is convenient to use these
as the independent hydrodynamic variables. This means that, apart from the
balance equations (\ref{2.8}) and (\ref{2.9}) for ${\bf u}$ and $T$, we need to get the corresponding balance equations for $x_{1}$ and $p$. These equations  can be easily obtained from (\ref{2.9}) 
and (\ref{2.11}) and are given by
\begin{equation}
D_{t}x_{1}+\frac{\rho }{n^{2}m_{1}m_{2}}\nabla \cdot {\bf j}_{1}=0\;,
\label{4.5}
\end{equation}
\begin{equation}
D_{t}p+p\nabla \cdot {\bf u}+\frac{2}{3}\left( \nabla \cdot {\bf q}+
{\sf P}:\nabla {\bf u}\right) =-\zeta p.  \label{4.6}
\end{equation}
Therefore, when the expressions of the fluxes (\ref{4.2})--(\ref{4.4}), 
and the cooling rate $\zeta \rightarrow \zeta ^{(0)}$ are
substituted into the balance equations (\ref{2.8}), (\ref{2.9}), (\ref{4.5}), 
and (\ref{4.6}) one gets a closed set of hydrodynamic equations for $
x_{1}$, ${\bf u}$, $T$, and $p$. These are the  Navier-Stokes equations for a mixture, which are
given by 
\begin{equation}
D_{t}x_{1}=\frac{\rho }{n^{2}m_{1}m_{2}}\nabla \cdot \left( \frac{m_{1}m_{2}n
}{\rho }D\nabla x_{1}+\frac{\rho }{p}D_{p}\nabla p+\frac{\rho }{T}D^{\prime
}\nabla T\right) \;,  \label{4.6a}
\end{equation}
\begin{eqnarray}
\left( D_{t}+\zeta \right) p+\frac{5}{3}p\nabla \cdot {\bf u} &=&\frac{2}{
3}\nabla \cdot \left( T^{2}D^{\prime \prime }\nabla x_{1}+L\nabla p+\lambda
\nabla T\right)  \nonumber \\
&&+\frac{2}{3}\eta \left( \nabla _{\ell }u_{k}+\nabla _{k}u_{\ell }-\frac{2}{
3}\delta _{k\ell }\nabla \cdot {\bf u}\right) \nabla _{\ell }u_{k},
\label{4.6b}
\end{eqnarray}
\begin{eqnarray}
\left( D_{t}+\zeta \right) T+\frac{5}{3}p\nabla \cdot {\bf u} &=&
-\frac{T}{n}\left( \frac{1}{m_{1}}-\frac{1}{m_{2}}\right) \nabla \cdot
\left( \frac{m_{1}m_{2}n}{\rho }D\nabla x_{1}+\frac{\rho }{p}D_{p}\nabla p+
\frac{\rho }{T}D^{\prime }\nabla T\right)  \nonumber \\
&&+\frac{2}{3n}\nabla \cdot \left( T^{2}D^{\prime \prime }\nabla
x_{1}+L\nabla p+\lambda \nabla T\right)  \nonumber \\
&&+\frac{2}{3n}\eta \left( \nabla _{\ell }u_{k}+\nabla _{k}u_{\ell }-\frac{2
}{3}\delta _{k\ell }\nabla \cdot \mathbf{u}\right) \nabla _{\ell }u_{k},
\label{4.6c}
\end{eqnarray}
\begin{equation}
D_{t}u_{\ell}+\nabla_{\ell}p=\rho ^{-1}\nabla _{k}\eta \left( \nabla _{\ell
}u_{k}+\nabla _{k}u_{\ell }-\frac{2}{3}\delta _{k\ell }\nabla \cdot {\bf u}\right) \;.
\label{4.6d}
\end{equation}
\begin{figure}
\includegraphics[width=0.6 \columnwidth]{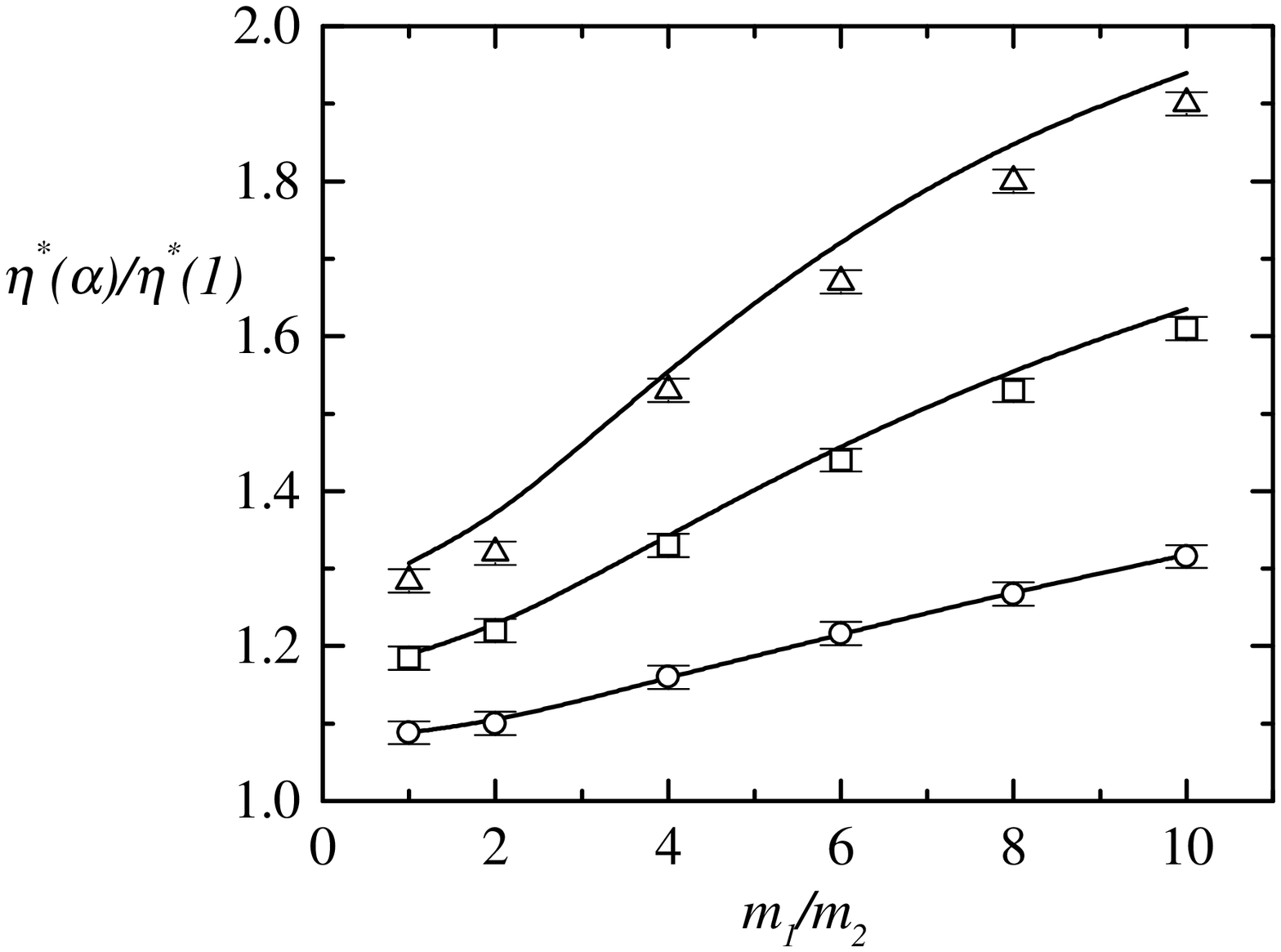}
\caption{Plot of the ratio $\eta^*(\alpha)/\eta^*(1)$ as a function of the
mass ratio $m_1/m_2$ for $\sigma_1/\sigma_2=n_1/n_2=1$ and three different
values of the restitution coefficient $\alpha$: $\alpha=0.9$ 
(circles), $\alpha=0.8$ (squares), and $\alpha=0.7$ (triangles).
The lines are the theoretical predictions and the symbols refer to the results
obtained from the DSMC method.
\label{fig3}}
\end{figure}
\begin{figure}
\includegraphics[width=0.6 \columnwidth]{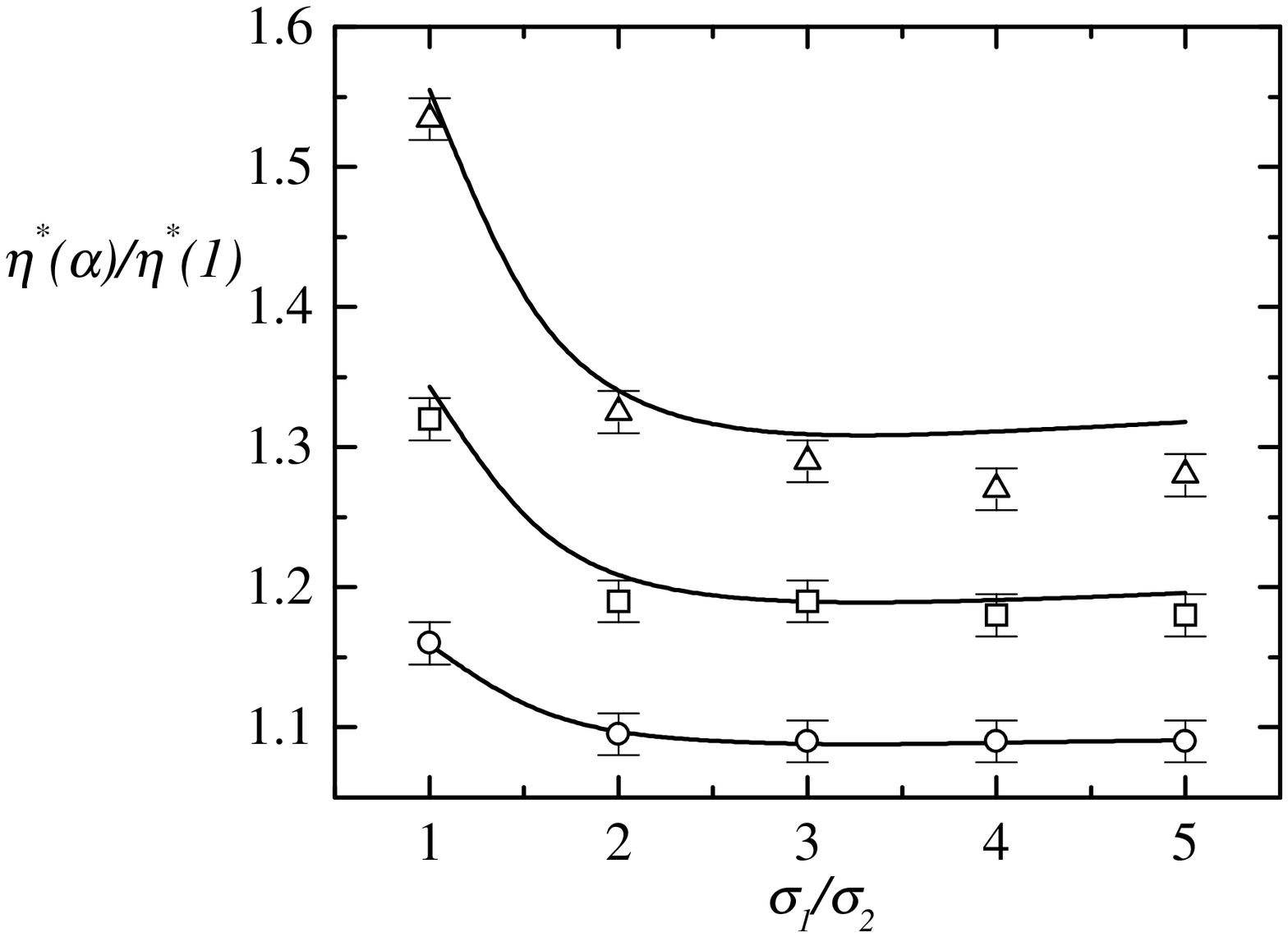}
\caption{Plot of the ratio $\eta^*(\alpha)/\eta^*(1)$ as a function of the
size ratio $\sigma_1/\sigma_2$ for $m_1/m_2=4$, $n_1/n_2=1$
and three different
values of the restitution coefficient $\alpha$:  $\alpha=0.9$ 
(circles),  $\alpha=0.8$ (squares), and $\alpha=0.7$ (triangles).
The lines are the theoretical predictions and the symbols refer to the results
obtained from the DSMC method.  
\label{fig4}}
\end{figure}

In general, the influence of dissipation on the transport coefficients is quite important even for moderate values of the restitution coefficient (say $\alpha\simeq 0.9$). Further, the effect of the additional contributions to the transport coefficients coming from the partial temperature differences can be quite significant, even for weak dissipation. These new contributions were not taken into account in previous work on granular mixtures\cite{Jenkins}.  The accuracy of the approximate Sonine solution to the resulting integral equations has been recently tested\cite{MG03}  at the level of the shear viscosity coefficient. Specifically, the DSMC method has been used to solve the Boltzmann equation in the special hydrodynamic state of uniform shear flow. An external thermostat is introduced to compensate for the energy lost in collisions so that the shearing work still heats the system. In the long time limit, the Navier-Stokes shear viscosity coefficient can be measured from simulations and compared with the predictions given by the Chapman-Enskog method. 
In Fig.\ \ref{fig3}, we plot the ratio $\eta^*(\alpha)/\eta^*(1)$ versus the mass ratio $m_1/m_2$ for $\sigma_1/\sigma_2=1$, $x_1=\case{1}{2}$, and three different values of the (common) restitution coefficient $\alpha$. Here, $\eta^*=\nu \eta/nT$, with $\nu=\sqrt{\pi}n\sigma_{12}^2v_0$ and $\eta^*(1)$ refers to the elastic value of the shear viscosity coefficient. Again, the symbols represent the simulation data while the lines refer to the theoretical results obtained from the Boltzmann equation in the first Sonine approximation. We see that 
in general the deviation of $\eta^*(\alpha)$ from its functional form for 
elastic collisions is quite important. This tendency becomes more 
significant as the mass disparity increases. The agreement between the first 
Sonine approximation and simulation is seen to be in general excellent. 
This agreement is similar to the one previously found in the monocomponent 
case\cite{BMC99,GM02}. At a quantitative level, the discrepancies between theory 
and simulation tend to increase as the restitution coefficient decreases, 
although these differences are quite small (say, for instance, around 2\% 
at $\alpha=0.7$ in the disparate mass case $m_1/m_2=10$). 
The influence of the size ratio on the shear viscosity is shown in Fig.\ 
\ref{fig4} for $m_1/m_2=4$ and $x_1=\case{1}{2}$. We observe again a strong 
dependence of the shear viscosity on dissipation. However, for a given value 
of $\alpha$, the influence of $\sigma_1/\sigma_2$ on $\eta^*$ is weaker than 
the one found before in Fig.\ \ref{fig3} for the mass ratio. 
The agreement for both $\alpha=0.9$ and $\alpha=0.8$ is quite good, except 
for the largest size ratio at $\alpha=0.8$. These discrepancies become more 
significant as the dissipation increases (say  $\alpha=0.7$),  
especially for mixtures of particles of very different sizes.    
In summary, according to the comparison carried out in Figs.\ \ref{fig3} and \ref{fig4}, one can conclude that the agreement between theory and simulation extends over a wide range 
values of the restitution coefficient, indicating the reliability of the first Sonine approximation for describing granular flows beyond the quasielastic limit.

\section{Linearized hydrodynamic equations and stability}
\label{sec5}

The simplest application of the hydrodynamic equations is to small
initial perturbations about the strictly homogeneous state for a large
system. Then the nonlinear equations can be expanded to linear order in the 
deviations of $x_{1}$, ${\bf u}$, $T$, and $p$ about their homogeneous values. This leads to partial differential equations with coefficients that are independent of space but they depend on time since the reference state is cooling. As in the monocomponent case\cite{BDKS98}, this time dependence can be eliminated through a change in the time and space variables, and a scaling of the hydrodynamic fields.  Let $\delta y_{\alpha}({\bf r},t)=y_{\alpha}({\bf r},t)-y_{H \alpha}(t)$ denote the deviation of $\{x_1, {\bf u}, T, p\}$ from their values in the homogeneous state. We introduce the following dimensionless space and time variables: 
\begin{equation}
\label{5.3}
\tau=\int_{0}^{t'} dt' \nu_{H}(t'),\quad {\boldsymbol {\ell}}=\frac{\nu_{H}(t)}{v_{0H}(t)}{\bf r},
\end{equation}
where $v_{0H}(t)$ and $\nu_H(t)$ has been introduced before. The superscript $H$ indicates that the quantity is evaluated in the homogeneous cooling state. This means that 
\begin{equation}
\label{5.4}
\partial_t x_{1H}=0, \quad {\bf u}_H={\bf 0}, \quad \partial_t \ln T_H=\partial_t \ln p_H=-\zeta_H
\end{equation}
A set of Fourier transformed dimensionless variables are then defined by 
\begin{equation}
\label{5.5}
\delta y_{{\bf k}\alpha}(\tau)=\int d{\boldsymbol {\ell}}\; e^{-i{\bf k}\cdot {\boldsymbol {\ell}}}\delta y_{\alpha}({\boldsymbol {\ell}},\tau),
\end{equation}
\begin{equation}
\label{5.6}
\rho_{{\bf k}}(\tau)=\frac{\delta x_{1{\bf k}}(\tau)}{x_{1H}}, \quad 
{\bf w}_{{\bf k}}(\tau)=\frac{\delta {\bf u}_{{\bf k}}(\tau)}{v_{0H}(\tau)},
\end{equation}
\begin{equation}
\label{5.7}
\theta_{{\bf k}}(\tau)=\frac{\delta T_{{\bf k}}(\tau)}{T_{H}(\tau)}, \quad 
\Pi_{{\bf k}}(\tau)=\frac{\delta p_{{\bf k}}(\tau)}{p_{H}(\tau)},
\end{equation}

In terms of the above variables the linearized hydrodynamic equations for the set $\{ \rho_{{\bf k}}, {\bf w}_{{\bf k}}, \theta_{{\bf k}}, \Pi_{{\bf k}} \}$ separate into a degenerate pair of
equations for the transverse velocity components $w_{\mathbf{k}\perp}$ 
(orthogonal to ${\bf k}$)  
\begin{equation}
\left( \frac{\partial }{\partial \tau }-\frac{\zeta ^{*}}{2}+\eta ^{*
}k^{2}\right) w_{{\bf k}\perp}=0,  \label{5.10}
\end{equation}
and a coupled set of equations for $\rho _{{\bf k}},\theta _{
{\bf k}},\Pi _{{\bf k}}$,  and the longitudinal \ velocity
component $w_{{\bf k}||}$ (parallel to ${\bf k}$) 
\begin{equation}
\frac{\partial \delta y_{{\bf k}\alpha }(\tau )}{\partial \tau }=\left(
M_{\alpha \beta }^{(0)}+ikM_{\alpha \beta }^{(1)}+k^{2}M_{\alpha \beta
}^{(2)}\right) \delta y_{{\bf k}\beta }(\tau ), 
\label{5.10a}
\end{equation}
where now $\delta y_{{\bf k}\alpha }(\tau )$ denotes
the four variables $\left( \rho _{{\bf k}},\theta _{{\bf k}},\Pi _{
{\bf k}},w_{{\bf k}||}\right)$. The matrices in this equation are
\begin{equation}
M^{(0)}=\left( 
\begin{array}{cccc}
0 & 0 & 0 & 0 \\ 
-x_{1}\left( \frac{\partial \zeta ^*}{\partial x_{1}}\right) _{T,p} & 
\frac{\zeta ^*}{2} & -\zeta ^{*} & 0 \\ 
-x_{1}\left( \frac{\partial \zeta ^{\ast }}{\partial x_{1}}\right) _{T,p} & 
\frac{\zeta ^{*}}{2} & -\zeta ^{*} & 0 \\ 
0 & 0 & 0 & \frac{\zeta ^{*}}{2}
\end{array}
\right),   \label{5.10b}
\end{equation}
\begin{equation}
M^{(1)}=\left( 
\begin{array}{cccc}
0 & 0 & 0 & 0 \\ 
0 & 0 & 0 & -\frac{2}{3} \\ 
0 & 0 & 0 & -\frac{5}{3}\\ 
0 & 0 & \frac{-\mu_{21}}{x_{1}\mu +x_{2}} & 0
\end{array}
\right),   \label{5.10c}
\end{equation}
\begin{equation}
M^{(2)}=\left( 
\begin{array}{cccc}
-D^{*} & -x_{1}^{-1}D^{'*} & -x_{1}^{-1}D_{p}^{\ast } & 0\\ 
-x_{1}\left( \frac{2}{3}D^{''*}-\frac{1-\mu }{x_{1}\mu
+x_{2}}D^{*}\right)  & \frac{1-\mu }{x_{1}\mu +x_{2}}D_{T}^{*}-\frac{
2}{3}\lambda ^{*} & -\frac{2}{3}L^{*}+\frac{1-\mu }{x_{1}\mu +x_{2}}
D_{p}^{*} & 0 \\ 
-\frac{2}{3}x_{1}D^{''*} & -\frac{2}{3}\lambda ^{*} & 
-\frac{2}{3}L^{*} & 0 \\ 
0 & 0 & 0 & -\frac{4}{3}\eta ^{*}
\end{array}
\right).   \label{5.10d}
\end{equation}
In these equations, $\mu =m_{1}/m_{2}$, $x_{i}=n_{iH}/n_{H}$, $\zeta ^{*
}=\zeta _{H}^{(0)}/\nu _{H}$.  Moreover, we have introduced the reduced
Navier-Stokes transport coefficients 
\begin{equation}
\label{5.13}
D^*=\frac{\nu_H D}{n_Hv_{0H}^2},\quad D_p^*=\frac{\rho_H^2\nu_H D_p}{m_1m_2n_H^2v_{0H}^2},\quad 
D^{'*}=\frac{\rho_H^2\nu_H D'}{m_1m_2n_H^2v_{0H}^2},
\end{equation}
\begin{equation}
\label{5.14}
\eta^*=\frac{\nu_H \eta}{\rho_H v_{0H}^2},
\end{equation}
\begin{equation}
\label{5.15}
D^{''*}=\frac{\nu_H T_HD''}{n_Hv_{0H}^2},\quad L^*=\frac{\nu_H L}{v_{0H}^2},\quad 
\lambda^{*}=\frac{\nu_H \lambda}{n_Hv_{0H}^2}.
\end{equation}

Equation (\ref{5.10}) is decoupled from the rest and can be integrated directly leading to 
\begin{equation}
\label{5.16}
w_{{\bf k}\perp}(\tau)=w_{{\bf k}\perp}(0)\exp(s_{\perp}\tau), 
\end{equation}
where 
\begin{equation}
\label{5.17}
s_{\perp}=\frac{1}{2}\zeta^*-\eta^* k^2.
\end{equation}
This identifies two shear modes. The remaining modes have the form $\exp(s_{n}\tau)$ for $n=1,2,3,4$, where $s_n$ are the solutions of the corresponding quartic equation. A detailed study on the dependence of the hydrodynamic modes on the inelasticity and the parameters of the mixture will be given elsewhere. The linear hydrodynamic equations characterize the stability of the homogeneous cooling state. We see from Eq.\ (\ref{5.17}) that $s_{\perp}$ for the two shear modes   is positive for $k<k_{\perp}^{\text{c}}$, where 
\begin{equation}
\label{5.18}
k_{\perp}^{\text{c}}=\left(\frac{\zeta^*}{2\eta^*}\right)^{1/2}.
\end{equation}
Thus, initial long wavelength perturbations of the HCS that for instance, excite the shear mode will grow exponentially, representing an instability of the reference state. 

The wave vector dependence of the remaining four modes is more
complex. However, the long wavelength stability can be obtained from the
eigenvalues of $M^{(0)}$ which are easily found to be
\begin{equation}
\label{5.n1}
s_{n}=\left(-\frac{1}{2}\zeta ^{*},0,0,\frac{1}{2}\zeta ^{*}\right).
\end{equation}
Hence, there is another unstable mode. In summary, at asymptotically
long wavelengths the spectrum of the linearized hydrodynamic equations is
comprised of a decaying mode at $-\zeta ^{*}/2$, a two-fold
degenerate mode at $0$, and a three-fold degenerate unstable mode
at $\zeta ^{*}/2$.

\section{Discussion}
\label{sec6}

The primary objective of this review has been to obtain the hydrodynamic description of a binary mixture of granular gases from the Boltzmann kinetic theory. The Chapman-Enskog method is used to solve the Boltzmann equation up to 
the Navier-Stokes order and the associated transport coefficients are given in terms of the solutions of a set of linear integral equations. A practical evaluation of these coefficients is possible by taking the leading terms in a Sonine polynomial expansion. In contrast to previous studies \cite{Jenkins}, our results apply for arbitrary degree of dissipation and they are not restricted to specific values of the parameters of the mixture. The explicit knowledge of the transport coefficients and the cooling rate allows one to perform an study  of the linearized hydrodynamic equations around the homogenous cooling state. The stability analysis shows that the homogeneous cooling state is unstable to long enough wavelength perturbations.

The reference state in our Chapman-Enskog expansion has been taken to be an exact solution of the uniform Boltzmann equation. An interesting and important result of this solution is that the partial temperatures of each species must be different. This does not mean that there are additional hydrodynamic degrees of freedom since their cooling rates are equal and consequently, the partial temperatures can be expressed in terms of the global temperature. However, the relationships between these partial temperatures are functions of the mole fractions (composition) and lead to new contributions (not considered in previous works) to the transport coefficients. The consequences of this new effect on transport are quite significant.

The dependence of the transport coefficients on the full parameter space has been explored in the case of the shear viscosity $\eta$. In addition, these theoretical predictions have been compared with the results obtained from a numerical solution of the Boltzmann equation by means of the DSMC method. The theory and simulation clearly show that in general, the influence of dissipation on $\eta$ is quite important since there is a relevant dependence of the shear viscosity on the restitution coefficient. With respect to the accuracy of the Chapman-Enskog results we see that the first Sonine approximation exhibits an excellent agreement with the simulation data. This supports the idea that the Sonine polynomial approximation for granular fluids has an accuracy comparable to that for elastic collisions. Exceptions to this agreement are extreme mass ratios and strong dissipation. These discrepancies are due basically to the approximations introduced in applying the Chapman-Enskog, and more specifically in using the first Sonine approximation.      

Although the utility of a hydrodynamic description for granular media under rapid flow conditions has been recognized for many years, there are some doubts about the possibility of going from a kinetic theory to a hydrodynamic level of description by using a Chapman-Enskog expansion around the homogeneous cooling state. Given that the search for exact solutions of the Boltzmann equation is far beyond the present perspectives, an alternative to get some insight into the above question is to numerically solve the kinetic equation by using for instance the DSMC method and compare these results with the corresponding solution obtained by assuming the validity of a hydrodynamic description. The good agreement obtained here for the temperature ratio and the shear viscosity coefficient shows clearly the direct verification of hydrodynamics and the quantitative predictions for transport coefficients from kinetic theory for states with small spatial gradients but including strong dissipation. 

One of the main limitations of the results obtained here from the Boltzmann equation is its restriction to the low-density regime. In this regime, the collisional transfer contributions to the fluxes are negligible and only their kinetic contributions are taken into account. Possible extension in both aspects, theory and simulation, of the present hydrodynamic description to higher 
densities can be carried out in the context of the revised Enskog theory. In this case, many of the phenomena appearing in dense granular fluids (such as spontaneous formation of 
dense clusters surrounded by regions of low-density \cite{GZ93}) could be studied.

The instability of the HCS found in the previous section may also
seem to represent a limitation on the validity of hydrodynamics. However, it
should be noted that the Chapman-Enskog expansion is not an expansion about
the HCS but rather its local form, parametrized by the exact hydrodynamic
fields. These fields obey the nonlinear hydrodynamic equations which are not
unstable, but rather they describe the correct subsequent evolution of the
unstable linear equations. The linearized hydrodynamics is indeed limited to
time scales short compared to the growth of the linear perturbations. Nevertheless, these equations describe correctly the nature of the onset of
the instability. Since it is a long wavelength phenomenon, the instability
can be suppressed completely for sufficiently small system sizes.

\acknowledgments  
V. G. acknowledges partial support from the Ministerio de Ciencia y
Tecnolog\'{\i}a (Spain) through Grant No. BFM2001-0718. The research of
J.W.D. was supported by Department of Energy Grants DE-FG03-98DP00218 and
DE-FG02ER54677.


\begin{references}


\bibitem{BDKS98} J. J. Brey, J. W. Dufty, C. S. Kim, and A. Santos, Phys. Rev. E {\bf 58}, 4638 (1998). 


\bibitem{BDS97}J. J. Brey, J. W. Dufty, and A. Santos, J. Stat. Phys. {\bf 87}, 1051 (1997).


\bibitem{CC70}S. Chapman and T. G. Cowling, {\em The Mathematical Theory of Nonuniform Gases} (Cambridge University Press, Cambridge, 1970).  


\bibitem{Jenkins}J. T. Jenkins and F. Mancini, Phys. Fluids A {\bf 1}, 2050 (1989); P. Zamankhan, Phys. Rev. E {\bf 52}, 4877 (1995); B. Arnarson and J. T. Willits, Phys. Fluids {\bf 10}, 1324 (1998); J. T. Willis and B. Arnarson, Phys. Fluids {\bf 11}, 3116 (1999); M. Alam, J. T. Willits, B. Arnarson, and S. Luding, Phys. Fluids {\bf 14}, 4085 (2002).

\bibitem{GD99}V. Garz\'o and J. W. Dufty, Phys. Rev. E {\bf 60}, 5706 (1999).

\bibitem{BT02}A. Barrat and E. Trizac, Granular Matter {\bf 4}, 57 (2002).

\bibitem{MP02}U. M. B. Marconi and A. Puglisi, Phys. Rev. E {\bf 65}, 051305 (2002); Phys. Rev. E {\bf 66}, 011301 (2002).


\bibitem{NK02}E. Ben-Naim and P. L. Krapivsky, Eur. Phys. J. E {\bf 8}, 507 (2002).

\bibitem{MG02}J. M. Montanero and V. Garz\'o, Granular Matter {\bf 4}, 17 (2002). 

\bibitem{DHGD02}S. R. Dahl, C. M. Hrenya, V. Garz\'o, and J. W. Dufty, Phys. Rev. E {\bf 66}, 041301 (2002). 


\bibitem{CH02}R. Clelland and C. M. Hrenya, Phys. Rev. E {\bf 65}, 031301 (2002). 

\bibitem{WP02}R. D. Wildman and D. J. Parker, Phys. Rev. Lett. {\bf 88}, 064301 (2002).


\bibitem{FM02}K. Feitosa and R. Menon, Phys. Rev. Lett. {\bf 88}, 198301 (2002).


\bibitem{GD02} V. Garz\'o and J. W. Dufty, Phys. Fluids {\bf 14}, 1476 (2002).



\bibitem{MG03}J. M. Montanero and V. Garz\'o, Phys. Rev. E {\bf 67}, 021308 (2003). 


\bibitem{B94}G. A. Bird, {\em Molecular Gas Dynamics and the Direct Simulation Monte Carlo of Gas Flows} (Clarendon, Oxford, 1994).


\bibitem{GS95}A. Goldhstein and M. Shapiro, J. Fluid Mech. {\bf 282}, 75 (1995). 


\bibitem{NE98}T. P. C. van Noije and M. H. Ernst, Granular Matter {\bf 1}, 57 (1998).  

\bibitem{GD99b}V. Garz\'o and J. W. Dufty, Phys. Rev. E {\bf 59}, 5895 (1999).

\bibitem{BMC99}J. J. Brey, M. J. Ruiz-Montero, and D. Cubero, Europhys. Lett. {\bf 48}, 359 (1999). 

\bibitem {GM02} V. Garz\'o and J. M. Montanero, Physica A {\bf 313}, 336 (2002).


\bibitem{GZ93} I. Goldhirsch and G. Zanetti, Phys. Rev. Lett. {\bf 70}, 1619 (1993).


\end{references}
\end{document}